\documentclass[12pt,dvips]{article}
\usepackage[dvips]{graphicx}
\usepackage{graphicx}
\usepackage{epsfig}
\usepackage{amsmath,amssymb,amsfonts}
\usepackage{cite}
\usepackage[x11names]{xcolor}

\parskip=\itemsep               
\newcommand{\lambdabar}{{\hbox{$\lambda_e$\kern-1.9ex\raise+0.45ex\hbox{--}
\kern+0.2ex}}}

\def\beq{\begin{equation}}
\def\eeq{\end{equation}}

\newif\ifhepph

\newcommand{\mg}{m_{\gamma^{\prime}}}
\newcommand{\muu}{m_{\gamma^{\prime}}}

\newcommand{\fin}{{\cal F}}

\renewcommand\({\left(}
\renewcommand\){\right)}
\renewcommand\[{\left[}

\newcommand{\be}{\begin{equation}}
\newcommand{\ee}{\end{equation}}
\def\bea{\begin{eqnarray}}
\def\eea{\end{eqnarray}}

\newcommand{\ff}{{\cal F}}

\ifhepph\date{\empty}\fi

\title{
{\normalsize \rightline{DESY 09-037; IPPP/09/22; DCPT/09/44}}\ \vskip 1cm \bf\boldmath
Hidden Laser Communications Through Matter\\[1.5ex] --
An application of meV-scale hidden photons --
       \vspace{21mm}}

\author{
Joerg Jaeckel$^{1,}$\footnote{{\bf e-mail}: joerg.jaeckel@durham.ac.uk}\,\,,
Javier Redondo$^{2,}$\footnote{{\bf e-mail}: javier.redondo@desy.de}\,\,, and
Andreas Ringwald$^{2,}$\footnote{{\bf e-mail}: andreas.ringwald@desy.de}
\\[2ex]
\small{\em $^1$ Institute for Particle Physics and Phenomenology, Durham University, Durham DH1 3LE, UK}\\
\small{\em $^2$Deutsches Elektronen-Synchrotron, Notkestra\ss e 85, 22607 Hamburg, Germany}
}

\begin{document}
\begin{titlepage}
  \maketitle
\begin{abstract}
Currently, there are a number of light-shining-through-walls experiments searching for hidden photons -- light, sub-eV-scale,
abelian gauge bosons beyond the standard model which mix kinetically with the standard photon.
We show that in the case that one of these experiments finds
evidence for hidden photons, laser communications through matter, using
methods from free-space optics, can be realized in the very near future, with a channel
capacity of more than 1 bit per second, for a distance up to the Earth's diamater.
\end{abstract}


\thispagestyle{empty}
\end{titlepage}
\newpage \setcounter{page}{2}

\begin{figure}[t]
\centerline{\includegraphics[width=9cm]{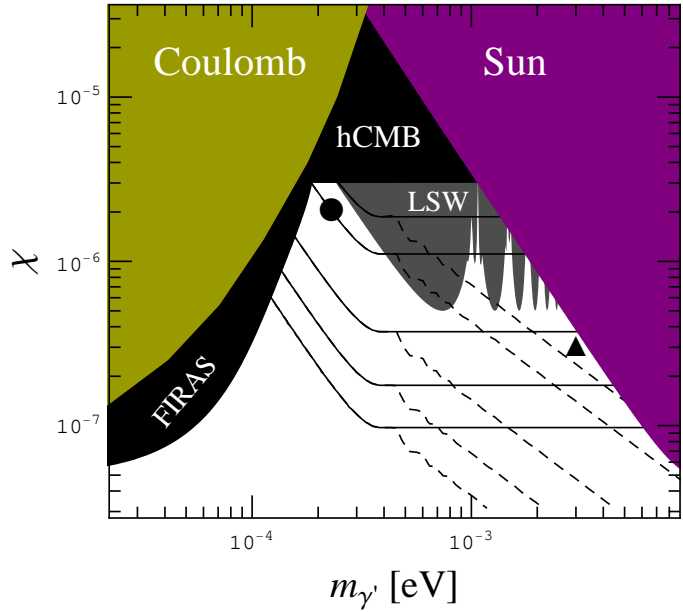}}
\caption{
Isocontours of channel capacity (bit/s) for our hidden photon communication system
across the Earth's diameter.
>From bottom to top the channel capacities are $1,10,100,500,1000$ bit/s.
The dashed (solid) lines correspond to setups with (without) phase shift plates.
Further details are given in the main text.
The dot and triangle are our benchmark points $(\muu^\bullet,\chi^\bullet)$ and $(\muu^\blacktriangle,\chi^\blacktriangle)$.
Also shown are current experimental limits on the possible existence of a hidden photon of
mass $m_{\gamma^\prime}$, mixing kinetically with the photon, with a mixing parameter
$\chi$. Strong constraints arise from the non-observation of deviations from
the Coulomb law (yellow)~\cite{Williams:1971ms,Bartlett:1988yy},
from Cosmic Microwave Background (CMB) measurements of the effective number of neutrinos and the blackbody nature of the spectrum (black)~\cite{Jaeckel:2008fi}, from light-shining-through-walls (LSW) experiments (grey)~\cite{Ruoso:1992nx,Cameron:1993mr,Robilliard:2007bq,Ahlers:2007rd,Chou:2007zzc,Ahlers:2007qf,Afanasev:2008jt,Fouche:2008jk,Afanasev:2008fv}, and from
searches of solar hidden photons with the CAST experiment (purple)~\cite{Andriamonje:2007ew,Redondo:2008aa}.
The white region in parameter space is currently unexplored, but may be accessed by
experiments in the very near future, in particular by improvements in LSW experiments (for proposed experiments probing
this region, see Refs.~\cite{Jaeckel:2007ch,Gninenko:2008pz,Jaeckel:2008sz}).
 }\label{Fig:current_limits}
\end{figure}

Many extensions of the standard model predict one or more new abelian gauge bosons ($\gamma^\prime$) besides the photon
($\gamma$).
If they are massless or very light, with masses $m_{\gamma^\prime}$ in the sub-eV range, their dominant interaction with
the standard photon arises from a mixing in the gauge-kinetic terms in the Lagrangian~\cite{Holdom:1985ag},
\begin{equation}
\label{lagrangian}
{\mathcal{L}}= -\frac{1}{4} F^{\mu\nu}F_{\mu\nu}-\frac{1}{4}B^{\mu\nu}B_{\mu\nu}
-\frac{1}{2}\chi\,F^{\mu\nu}B_{\mu\nu}  +\frac{1}{2}\mg^2 B_\mu B^\mu+j_\mu A^\mu,
\end{equation}
where $F_{\mu\nu}$ is the field strength tensor for the ordinary
electromagnetic gauge field $A^{\mu}$, $j^\mu$ is its associated current (generated by
electrons, etc.),  and $B^{\mu\nu}$ is the
field strength for the new abelian gauge field
$B^{\mu}$. The parameter $\chi$ in Eq.~(\ref{lagrangian}) gives the strength of the
kinetic mixing between $A$ and $B$.
Within the context of string-inspired extensions of the standard model, it is expected to lie in the range between $10^{-23}$ and $10^{-2}$ (cf.~\cite{Dienes:1996zr,Abel:2003ue,Abel:2006qt,Abel:2008ai}), while experimentally or phenomenologically,
the current limits, in the micro-eV up to eV range, are displayed in Fig.~\ref{Fig:current_limits}.

A prominent role in these current limits are played by laboratory experiments exploiting the
light-shining-through-walls (LSW) technique~\cite{Okun:1982xi,Anselm:1986gz,Anselm:1987vj,VanBibber:1987rq,Pugnat:2007nu,Ehret:2008sj}.
In these experiments, laser light is shone through a vacuum tube and blocked from another vacuum tube that is aligned to the first
one. Hidden photons eventually generated in the first vacuum tube by photon -- hidden photon oscillations, induced by
kinetic mixing, will fly through the beam stopper due to their negligible interaction with matter and reconvert into
photons in the second vacuum tube, appearing as light shining through the wall.
In this letter, we show that in the case that one of the current experiments finds
evidence for hidden photons, laser communications through matter, using
methods from free-space optics, can be realized in the very near future, with a channel
capacity of more than 1 bit per second, for a distance up to the Earth's diameter.
A similar proposal has been recently discussed in~\cite{Stancil:2007yk} in which the
particles used for communication are not hidden photons but axion-like-particles (ALPs).
Unfortunately ALPs with the parameters required in~\cite{Stancil:2007yk} are excluded
by astrophysical considerations that seem quite difficult to evade (however, see~\cite{Masso:2006gc}).
In contrast, the hidden photons considered in this paper do not suffer from such problems.

\begin{figure}[th]
\centerline{\includegraphics[width=\textwidth]{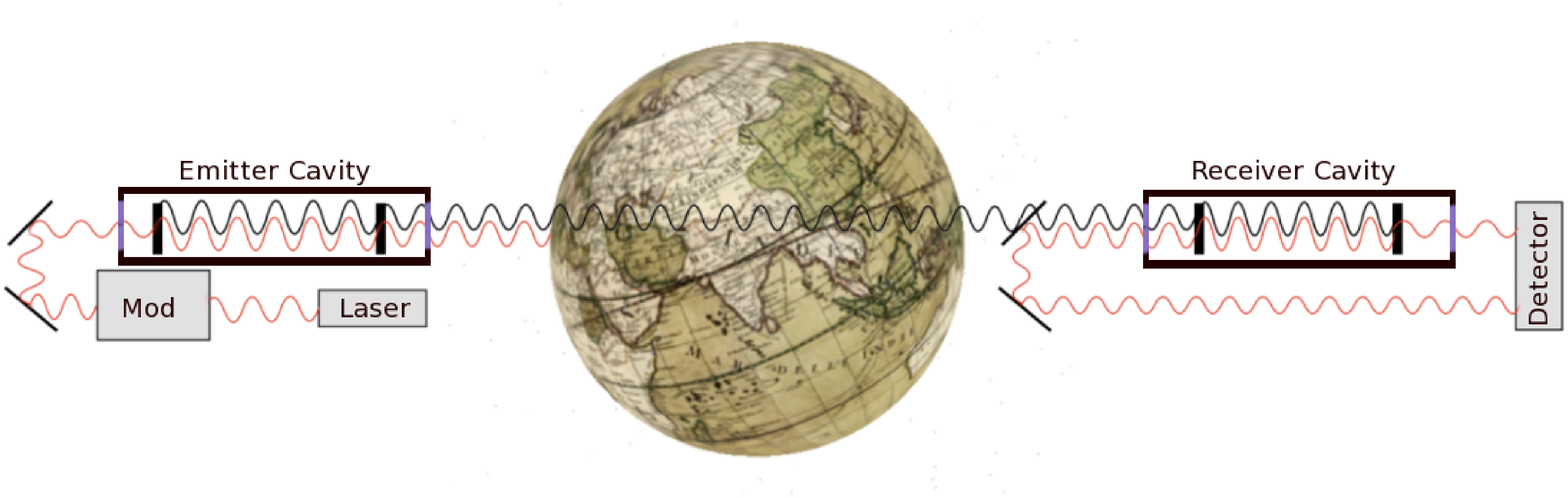}}
\caption{Sketch of the HP communication system. A laser modulated in
amplitude or polarization feeds a resonant cavity placed inside a
vacuum tube. A fraction of the power inside the cavity oscillates
into hidden photons that being weakly interacting scape the cavity
and can traverse dense media without loosing information. The
receiver is another cavity in vacuum locked to the frequency of the
HP signal. HPs resonantly reconvert into photons that are finally
detected at both the receiver cavity ends.  }\label{Fig:komm}
\end{figure}

The proposed communication system is based on the concept of free space optics, exchanging the role of optical light as the transmitter
for a beam of hidden photons (cf. Fig.~\ref{Fig:komm}).
Such a beam can be produced and controlled by a ``progenitor'' laser beam propagating in vacuum.
The probability of vacuum photon $\to$ hidden photon oscillations (and viceversa) is given by~\cite{Ahlers:2007rd}
\be
\label{prob}
P(\gamma\to \gamma^\prime ) =P(\gamma^\prime \to \gamma) = 4\chi^2 \sin^2\frac{\muu^2 L}{4 \omega} \equiv 4 \chi^2 \times a ,
\ee
with $\omega$ the laser photon frequency. The oscillation length is given by $L_{\rm osc}=4\pi\omega/\muu^2$.
Here we are interested in values in the meV valley (see Fig.~\ref{Fig:current_limits}). For concreteness we will consider the
following benchmark points
\bea
\label{benchp}
\muu^\bullet &=& 2.3 \times 10^{-4}\ {\rm eV}\quad ;\quad \chi^\bullet=2\times 10^{-6},     \\
\muu^\blacktriangle &=& 3.0\times 10^{-3}\ {\rm eV}\quad ;\quad \chi^\blacktriangle=3\times 10^{-7},
\eea
which can be probed by laboratory experiments in the near future
(cf. Fig.~\ref{Fig:current_limits}). In addition, the first one has
interesting phenomenological consequences in
cosmology~\cite{Jaeckel:2008fi}, while the second one could be
relevant in astrophysical contexts~\cite{Gninenko:2008pz}. The
maximum of the oscillation (for which $a=1$) is reached after a
length
\be
 L_{\rm osc}/2=23\ {\rm m} \left( m^\bullet/\muu \right)^{2}\left( \omega/{\rm eV } \right)=
 0.14\ {\rm m} \left( m^\blacktriangle/\muu \right)^{2}\left( \omega/{\rm eV } \right)\ .
\ee

Our system is based on two optical cavities%
\footnote{Increasing the sensitivity of LSW experiments by putting
resonant optical cavities on both sides of the wall was first proposed in Ref.~\cite{Hoogeveen:1990vq} and
rediscovered in Ref.~\cite{Sikivie:2007qm}.}
which act as $\gamma\leftrightarrow \gamma^\prime$ transducers:
emitter and receiver, cf.~Fig.~\ref{Fig:komm}.
In the emitter, hidden photons are produced through $\gamma\to\gamma^\prime$ oscillations with the same characteristics as the photons in the cavity and therefore their beam can be easily controlled.
A modulator (Mod) can actuate on the laser amplitude, phase, or polarization, the cavity being a mere amplification mechanism.

The emitted hidden photons can traverse any dense medium without significant losses. The hidden photon absorption length in a dense medium with photon
absorption length $l_\gamma$ can be estimated as follows (see, for instance~\cite{Redondo:2008aa})
\be
l_{\gamma^{\prime}}\geq l_\gamma
\frac{L_{\rm osc}^2}{\chi^2 l_\gamma^2} =
5.3\times 10^{9}\, {\rm km}
\(\frac{l_\gamma}{1\, \rm mm}\)^{-1}
\(\frac{\omega}{1\, \rm eV}\)
\(\frac{\chi}{\chi^\blacktriangle}\)^{-2}
\(\frac{\muu}{\muu^\blacktriangle}\)^{-4} \ .
\ee
For the $\bullet$ benchmark point (and in general the rest of the interesting parameter space) this is even longer.
Therefore, for the values we consider these losses are completely negligible, even in a dense medium like the Earth.

The only losses to consider are then diffraction losses.
Assuming that the emitter cavity is locked in the fundamental Gaussian mode (TEM$_{00}$) and is optimized for cavity mirrors of diameter $D_e$, the hidden photon power at a distance $R$ from the emitter is diminished by the factor
\be
\frac{\omega^2 D_e^2D_r^2}{8 \pi^2 R^2} \ .
\ee
This is the case when the input beam has a waist size that almost fills the mirror diameter $w=0.45 D_e$ (see for instance~\cite{Robertson:1996cx}).
Here $D_r$ is the diameter of the mirrors of the receiver cavity.

\begin{table*}[t]
\centering
\begin{tabular}{c|c|ccc}
Benchmark  & conf.  $\backslash$ R [km]   & $10^3$  & $12.8\times 10^3$ (D$_{\rm Earth}$)    &   384 $\times 10^3\ $ (R$_{\rm moon}$) \\
\hline\hline \\[-2ex]
$\bullet$ & 2 cav.                   &   2660 $|_{\ff = 2.3\times 10^4}$    &     486$|_{\ff =1.3\times 10^5}$  &   16$|_{\ff = 3.0\times 10^5}$   \\[0.2ex]
$(n_{\rm psp}=0)$  & 1 cav. &   23$|_{\ff = 3.0\times 10^5}$    &     0.16$|_{\ff = 3.0\times 10^5}$  &    1.8$\times 10^{-4} |_{\ff = 3.0\times 10^5}$   \\[0.2ex]
\hline\hline\\[-2ex]
$\blacktriangle$       & 2 cav.                   &   7440$|_{\ff = 8.2\ 10^3}$    &     1360$|_{\ff =4.5\ 10^4}$  &    132$|_{\ff =3.0\ 10^5}$   \\[0.2ex]
$(n_{\rm psp}=62)$  & 1 cav. &   1200$|_{\ff = 4.6\times 10^4}$    &    102 $|_{\ff = 3.0\times 10^5}$  &    0.35$|_{\ff = 3.0\times 10^5}$   \\[0.2ex]
\end{tabular}
\caption{Capacity in bit/s of hidden photon communication at different distances for hidden photon parameters
 given by the two benchmark points $\bullet,\blacktriangle$ (see Eq.~\eqref{benchp} and Fig.~\ref{Fig:current_limits}).
We show two different configurations: 2 cavities (at emitter and receiver) and 1 cavity (only at emitter).
The finesse of the cavities was optimized as described in the text with a maximum value of ${\cal F}=3\times 10^{5}$.
The optimized value of the finesse is indicated in the subscripts.
A number $n_{\rm psp}$ of phase shift plates has been used.}
\label{tab:cap}
\end{table*}

At the receiver tube, the hidden photon beam can be considered as a plane wave.
Assuming perfect alignment of the receiver cavity, the photon signal will be resonantly enhanced due to constructive interference of photons coming from oscillations of hidden photons that enter the cavity at different times.
The locking of this cavity has to be done to a fixed reference frequency or an atomic clock.
Photons in the receiver will exit the cavity in both directions and can be collected by two detectors.
The final power ${\rm P}_{\gamma,r}$ is then related to the input laser power ${\rm P}_{\gamma, 0}$ by the expression
\be {\rm P}_{\gamma,r} = 32 \chi^4  a_e a_r
\frac{\fin_e}{\pi}\frac{\fin_r}{\pi} \frac{\omega^2 D_e^2 D_r^2}{8
\pi^2 R^2}  \,{\rm P}_{\gamma, 0} ,
\ee
where $\fin_{e,r}$ are the finesses of the emitter and receiver cavities.

The channel capacity, i.e. the theoretical maximum of information that can be reliably transmitted over a communication channel, will generally depend on the process of detection of the photons exiting the receiver cavity.
Two schemes seems possible, direct photon detection or heterodyne amplification%
\footnote{This method is disfavored in purely optical communications due to the distortions caused by the atmosphere, but this is certainly not a limitation in our case.}.
In practice both methods should give similar results for the small signals we are considering.
In the latter case, the channel capacity in bit/s is~\cite{Chen:2006??}
\be
C=\Delta \nu \log_2 \left(1+\frac{S}{\Delta \nu}\right),
\ee
with $\Delta \nu$ the smaller bandwidth (in Hz) of the two
cavities ($\Delta \nu = (2 L {\cal F})^{-1}$), $S=\eta {\rm
P}_\gamma/\omega$ the number of photons detected with $\eta$ the
quantum efficiency.
As fiducial values we can take: $L\leq 10$ m,
$\omega=1.16$ eV ($\lambda = 1064$~nm), $\eta=0.86$, $D_{e,r}=38$ cm and
P$_{\gamma 0}=100$ W, values certainly attainable today.

Since $C$ is a convex function of $\cal F$, it can be easily maximized.
Writing $C=c_1{\cal F}^{-1}\log_2\(1+c_2{\cal F}^3\)$ we find an optimum for ${\cal F}=2.51 (c_2)^{-1/3}$ where formally $c_2=|S/\Delta \nu |_{{\cal F}=1}$.
Allowing values up to ${\cal F} = 3\times 10^5$ and our benchmark point ($\chi^\bullet,\muu^\bullet$), we
find the achievable channel capacities at different distances shown in Table~\ref{tab:cap}. In
Fig.~\ref{Fig:current_limits} we show the dependence on $\muu$ and
$\chi$ for a fixed distance taken to be the Earth's diameter.
Choosing a symmetric setup with two equal cavities allows for communication
in both directions: the ``receiver" cavity can also be fed by a laser and
the ``emitter" cavity can also be equipped with a detector system.
A more simple set-up can be achieved not including the receiver cavity.
Of course, the capacity in this case is much smaller, as also
shown in Tab.~\ref{tab:cap}.

This setup based on resonant cavities seems to be optimal for small masses like $\muu^\bullet=2.3\times 10^{-4}$ eV, but it can be improved in
the case of larger masses by the use of \emph{phase shift plates} (PSP)~\cite{Jaeckel:2007gk} inside
of the emitter and/or the receiver cavity.
Phase shift plates are thin small refractive plates whose optical path is tuned to restore the coherence of
the photon-hidden photon oscillations at the specific positions where it starts to decrease (this is a nearly optimal choice for the
number and location of the phase shift plates). Placing $n-1$ phase shift
plates with a spacing $L_{\rm osc}/2$ in an oscillation cavity the probability of the
oscillations after a length $n L_{\rm osc}/2$ is $4n^2\chi^2$ and therefore in principle the power transmitted would be
enhanced by a factor $n^4$. Unfortunately, the addition of optical components in the interior of a resonant optical cavity tends to decrease
the achievable finesse. For an impedance matched resonator~\cite{Lindner2009}, the finesse is typically inversely proportional to the dispersion
coefficient (in a round trip) which is directly proportional to the number of phase shift plates
\be
\frac{\cal F}{\pi}\simeq
\frac{4}{A}=\frac{4}{A_0+(n-1)A_{\rm PSP}}
\ee
with $A_0$ the loss factor due to absorption, scattering and deflection of light and transmissivity of the mirrors in the cavity without PSPs
and $A_{\rm psp}$ the loss factor of a PSP. For large $n$ the power will then  behave like $\sim n^2$.
For a fixed maximal length the number of phase shift plates is $2 L/L_{\rm osc}$.
For instance, in our second benchmark point with $\muu^\blacktriangle=3\times 10^{-3}$ eV and again using a cavity of length $10$ m this yields $n=62$ and an enhancement of the rate of order $\sim 4000$.
The capacities for several distances are shown in Tab.~\ref{tab:cap}.
The capacity improved by the insertion of phase shift plates is shown as dashed lines in Fig.~\ref{Fig:current_limits}.

{\em In summary,} we have proposed a method to send signals over long distances
through dense matter. Possible applications include communication between submarines, mines or to the backside of the moon~\cite{Stancil:2007yk}.
One could also contemplate the possibility of sending encryption keys directly through Earth thereby making it more difficult to eavesdrop.
Our system relies on the possible existence of a definite species of very light and very elusive particles beyond
the known ones: hidden photons. The latter may be emitted and received by
exploiting photon $\leftrightarrow$ hidden photon oscillations in resonant optical cavities. Because of their
very feeble interactions with known particles, the hidden photons emitted by the emitter cavity will not be
absorbed or deflected by any material between the latter and the receiver cavity.
Clearly, our proposal is an extreme version of an LSW experiment -- taking the wall thickness to extreme values.
With currently available
technology it seems that communication through the diameter of the Earth is possible with an information
transmission rate of more than 1 bit per second, provided hidden photons with sub-eV mass and
a kinetic mixing parameter, measuring the strength of the oscillations, exceeding $\chi \gtrsim 10^{-7}$ exist.
For masses in the $0.2-3$~meV range, this possibility is not only not excluded by experiments
(cf. Fig.~\ref{Fig:current_limits}), but it may even be expected theoretically.

Compared to a similar proposal~\cite{Stancil:2007yk}, which exploits axion-like-particles for the transmission, our proposal
has several advantages. First, the interesting parameter range for hidden photons is allowed by all current experiments and
observations. Second, the production and regeneration of hidden photon does not require a strong magnetic field making the
apparatus simpler. And finally the introduction of phase shift plates allows for a significant enhancement of the channel
capacity.

This provides further motivation for the current and near future small scale, precision optical experiments to probe the
meV mass range for hidden photons.

\section*{Acknowledgments}

We would like to thank A.~Lindner for encouragement and T.~Meier for discussions.

\end{document}